\documentclass[aps,prd,twocolumn,superscriptaddress,nofootinbib,10pt]{revtex4-2}

\usepackage[utf8]{inputenc}
\usepackage[T1]{fontenc}
\usepackage{lmodern}
\usepackage[english]{babel}
\usepackage{bm}
\usepackage{physics}
\usepackage{mathtools}

\usepackage{graphicx}
\usepackage{xcolor}
\usepackage{hyperref}
\usepackage{siunitx}
\usepackage{microtype}
\usepackage{enumitem}
\usepackage{amsmath,amssymb,amsfonts}
\usepackage[a4paper, total={6.9in, 10in}]{geometry}
\hypersetup{
  colorlinks = true,
  linkcolor  = blue,
  citecolor  = blue,
  urlcolor   = blue
}

\begin{document}

\title{Differential and Common Decoherence Modes in Witnessing the Quantum Gravity-Induced Entanglement of Matter}

\author{Ryan Rizaldy}
%\email{r.rizaldy@rug.nl}
\affiliation{Van Swinderen Institute for Particle Physics and Gravity, University of Groningen, 9747AG Groningen, the Netherlands}

\author{Helen M Sheehy}
\affiliation{Van Swinderen Institute for Particle Physics and Gravity, University of Groningen, 9747AG Groningen, the Netherlands}

\author{Tian Zhou}
\affiliation{Van Swinderen Institute for Particle Physics and Gravity, University of Groningen, 9747AG Groningen, the Netherlands}

\author{Anupam Mazumdar}
\affiliation{Van Swinderen Institute for Particle Physics and Gravity, University of Groningen, 9747AG Groningen, the Netherlands}
\affiliation{Canadian Institute of Theoretical Astrophysics, University of Toronto, Toronto, Canada}
%\date{\today}

\begin{abstract}
In the context of the QGEM (Quantum Gravity-induced Entanglement of Masses) experiment, we consider two adjacent matter-wave interferometers in linear and parallel configurations that interact solely via gravity. If gravity were quantum, then the two matter-wave interferometers would become entangled via the virtual excitation of the massless graviton. In this paper, we consider witnessing this entanglement by considering a generic experimental scenario where the two interferometers are subject to different global phases and different decoherence rates. In this context, we show that the individual global phases do not affect the witness, discuss common and differential decoherence modes, and perform the parameter search optimal for different masses. We provide a mathematical framework for these asymmetric decoherence rates and then search for parameters that determine the entanglement witness. We have kept the inter-separation distance between the two closest superpositions of the interferometers' masses fixed while varying the experimental time from $\tau=0.1$ s to $\tau=1$ s. Finishing the experiment at 
$ \tau=0.1$ s has many advantages from the point of view of protecting the experiment from random acceleration noise. However, witnessing the entanglement also suffers from $\langle W\rangle \sim -{\cal O}(10^{-2})$ for $m=10^{-14}$~kg, for 
decoherence rate in the ranges of ${\cal O}(10^{-1}-1)$~Hz for $\tau=0.1$ s experiment. However, as we show, increasing the mass of the matter-wave interferometer may improve the witness considerably.
\end{abstract}

\maketitle

% =====================================================================
\section{Introduction}
\label{sec:intro}

Testing the quantum nature of gravity is daunting because gravity is much weaker than the electromagnetic interaction. Despite this apparent weakness, we can observe the quantum nature of gravitational interaction via a protocol known as quantum gravity-induced entanglement of masses (QGEM). The protocol was first discussed in \cite{ICTS} and \cite{Bose:2017nin}; see also ~\cite{Marletto:2017kzi}, and relies on quantum correlations between incompatible (noncommuting) bases, known as entanglement. The gravitationally induced entanglement phase, which is nonvanishing if and only if gravity were quantum~\cite{Marshman:2019sne,Bose:2022uxe,Danielson:2021egj,christodoulou2023locally,Vinckers:2023grv,elahi2023probing,Carney_2019,Carney:2021vvt,Biswas:2022qto,Biswas:2026pem} can be witnessed in the QGEM protocol. A 
classical gravitational mediator will not generate entanglement due to the local operation and classical communication (LOCC) theorem~\cite{Bennett:1996gf}, as shown in \cite{Marshman:2019sne}, and an action at a distance will not yield entanglement because quantum mechanics does not allow instant interaction to develop due to the energy-time uncertainty relationship. A key ingredient of this experiment is creating a large spatial superposition for a nanoparticle of mass $m\sim 10^{-14}$~kg. In the original proposal, the authors proposed creating a superposition with $\Delta x\sim {\cal O }(250){\rm \mu m}$, the inter-separation distance between the two centres of mass to be around $d \sim {\cal O}(500)~{\rm \mu m}$, and keeping the superposition alive for roughly $\tau \sim 2.5$~s~\cite{Bose:2017nin}. However, some of these stringent conditions have been mitigated by introducing electromagnetic shielding \cite{vandeKamp:2020rqh,Schut:2023eux,Schut:2023hsy}, which enables witnessing entanglement for smaller superposition sizes $\Delta x \sim {\cal O}(5-50)~{\rm \mu m}$, and a separation distance around $d\sim 35~{\rm \mu m}$, while maintaining the experiment time of the order of $\tau \sim 1$ s.

However, witnessing entanglement requires one key ingredient in any quantum experiment: the decoherence rate,~\cite{Schlosshauer:2019ewh,bassireview,Romero-Isart:2011yun,RomeroIsart2011LargeQS,Hornberger_2012}, or, more precisely, how well we can maintain the coherence of the spatial superposition. For $\tau=1$s experiment, the decoherence rate must be $\gamma < 0.1$Hz. For the QGEM experiment, standard decoherence analysis has been done in \cite{Bose:2017nin,vandeKamp:2020rqh,Rijavec:2020qxd,Schut:2021svd,Schut:2023eux}. These are standard decoherence calculations based on collisional and blackbody emission and absorption. However, for the QGEM, additional decoherence sources arise from experimental systematics. For diamagnetic levitation of a nanodiamond, additional decoherence sources arise from ambient magnetic-field and current fluctuations~\cite{Moorthy:2025bpz, Fragolino:2023agd,Schut:2023tce}. Also, there is relative acceleration noise and Newtonian noise~\cite{Toros:2020krn,Wu:2024bzd,Wu:2024bzd}.

Hence, for a more practical solution, it is wise to use the entanglement witness. This is an observable or measurable variable (call it $W$) that has a special mathematical property: First, if the wavefunctions of the two spatial interferometers are not entangled (separable), the measured expectation value is always non-‐negative: $\langle W \rangle \geq 0$. Second, if the particles are entangled, the expectation value becomes negative: $\langle W \rangle < 0$. In particular, this is known as the positive partial transpose (PPT) witness~\cite{Horodecki:2009zz}, first used in the context of the QGEM experiment by \cite{Chevalier:2020uvv,Tilly:2021qef}.

 The QGEM experiment with nanodiamond relies on creating the spatial superposition via the Stern Gerlach interferometric scheme, for a review based on atoms, see~\cite{amit2019t,Folman2013,Folman2018,SGI_experiment}, and for recent discssion related to nanodiamond~\cite{Wan16_GM,WanPRA16_GM,Bose:2017nin,Pedernales:2020nmf,Marshman:2021wyk,Zhou:2022frl,Zhou:2022jug}. In the literature, there are two popular schemes: linear~\cite{Bose:2017nin} and parallel~\cite{Nguyen:2019huk} setups. 
 This paper studies both setups, including different global phases for the individual interferometers and different decoherence rates~\footnote{Previously, we analysed a specific case of parameter search in which we assumed the decoherence rates were the same for both interferometers~\cite{vandeKamp:2020rqh, Schut:2023hsy,Tilly:2021qef,Rijavec:2020qxd,toros.PRR2021,
 Schut:2025blz}. In general, they will differ between interferometers. The inclusive analysis should consider all of them separately for individual interferometers. }, and we make the following new claims:

\begin{itemize}

\item In all our previous studies, we assumed that the global phases of individual interferometers are the same; see~\cite{vandeKamp:2020rqh, Tilly:2021qef,Schut:2021svd,Schut:2023eux,Schut:2023hsy}. The latter would account for the phase induced by the conducting screen, or by any surface that induces Casimir or other electromagnetic interactions between the matter-wave interferometer and the nearby surfaces (including the wires). In general, these phases need not be the same. Hence, we treat these phases differently and keep our analysis general. We will explicitly show that the global phases from the two matter-wave interferometers simply add up over the experiment's integration time. Furthermore, they do not affect the entanglement witness, which is a seminal result for the QGEM experiment. Only the random decoherence rates affect the PPT witness.
%the entanglement phase generated between the two matter-wave interferometers should be larger than the sum of the global phases.
 
\item For the first time, we consider common and differential decoherence modes and how they affect the witness for different superposition sizes and different experimental times. In this regard, our analysis is model-independent; we constrain the full decoherence rates rather than individual noise sources that decohere the interferometers. 
We will vary the two decoherence rates for individual interferometers, denoted here by $\gamma_1$ and $ \gamma_2$. The case where they are equal will be the same as in the previous analysis \cite{Schut:2025blz}. However, allowing them to differ lets us better explore the experimental constraints of individual matter-wave interferometers and opens a larger parameter space, such as one interferometer being more decohered than the other, while still witnessing entanglement due to gravity.  In fact, our formalism is general and can be applied to perform a PPT witness for any interaction besides gravity, e.g., phonon- or photon-mediated entanglement~\footnote{We are clearly distinguishing global phases vs. the decoherence phase. Global phases are known classical phases, whereas the decoherence phase and its rate are not. Furthermore, we have an entanglement phase due to the quantum gravitational interaction between the two matter-wave interferometers.}.

\end{itemize}

 We analyse two time scales, $\tau=0.1$~s and 
 $\tau=1$ s. As we will see, finishing the experiment in $0.1$ s poses challenges; we will require a larger mass, i.e. $m=10^{-13}$~kg, to obtain a better entanglement witness for the decoherence rate $\gamma < 1$~Hz. Since the entanglement phase is $\propto m^2\tau$, increasing the mass by one order of magnitude is sufficient to obtain a better witness at $5-10~{\rm \mu m}$ spatial superposition. We adopt smaller timescales to finish the experiment to mitigate the random-acceleration noise limit obtained in ~\cite{toros.PRR2021, Bose:2026dag}. Also, for the $ t=0.1$ s experiment ($10$~Hz), the Einstein Telescope~\cite{hild2011sensitivity,ET:2025xjr,ETD_sensitivity_2018} can reach the random-acceleration noise limit required for the QGEM experiment, thereby protecting the experiment's coherence~\cite {Bose:2026dag}. 
 
% At $10$~Hz the strain noise is lower than $1$~Hz for the Einstein Telescope. Furthermore, this is a repetition experiment to gather statistics, and a longer-than-$1$ s experiment penalises us severely in terms of the total number of runs needed to infer a high percentage confidence level in witnessing the entanglement~\cite{Tilly:2021qef,Schut:2021svd}.

% In fact, there are many sources of decoherence, which would include collisions of air molecules with the nanoparticle of the matter-wave interferometer, blackbody emission and absorption \cite{Romero-Isart:2011yun}, induced and permanent dipole moment of the nanoparticle with the ambient air molecules~\cite{Fragolino:2023agd}, external electromagnetic jitters~\cite{Schut:2023tce}, current and magnetic field fluctuations~\cite{Moorthy:2025bpz}, decoherence induced by the conducting plate~\cite{Schut:2023eux}, NV spin and its interactions either with the impurities or the surface dangling bonds~\cite{WoodPRA22_GM}, and many unknown and perhaps unaccounted sources. 

 We begin with the simplest discussion of identical decoherence and identical global phases, then move to the main goals of this paper: accommodating the generic setup with different decoherence rates and global phases. 
 % =====================================================================
\section{Entanglement witness in the QGEM experiment}
\label{sec:witness}
% =====================================================================
\subsection{Pure state with identical decoherence between the two systems}
\label{sec:identical}

\begin{figure*}
    \centering
    \includegraphics[width=0.7\linewidth]{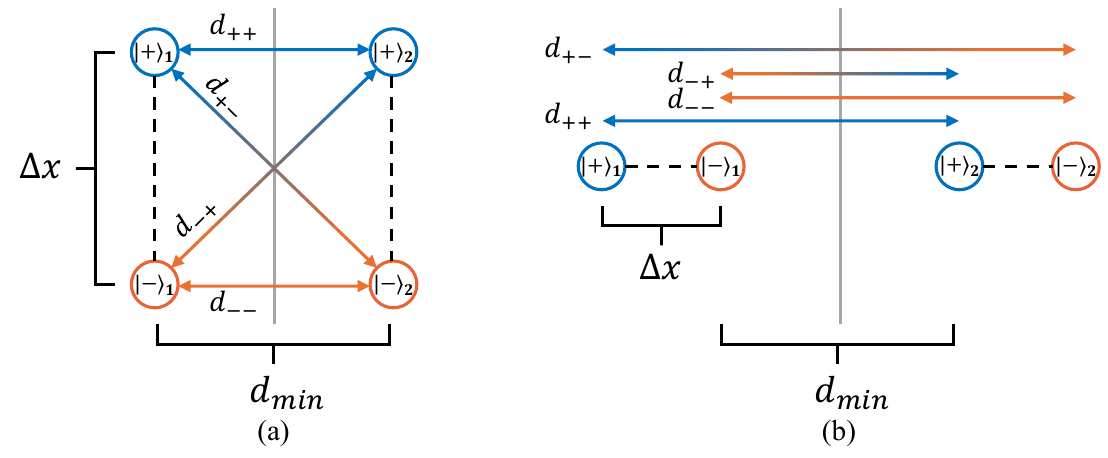}
    \caption{(a)Parallel~\cite{Nguyen:2019huk} and (b) linear setup~\cite{Bose:2017nin} of two test masses in spatial superpositions (1) and (2). The superposition sizes are assumed to be the same for both cases, $\Delta x$. We assume the internal spin states are split into $|+\rangle$ and $|-\rangle$ states. The spin-dependent Stern-Gerlach force creates the macroscopic spatial quantum superposition.
    The matter-wave interferometers are separated by a divider, shown here as a vertical line, which serves as an electromagnetic shield. The shield is treated classically here. The details can be found in QGEM references; see~\cite{vandeKamp:2020rqh,Schut:2023hsy}.}
    \label{fig:classicaltrajectoryHarmonic}
\end{figure*}

First, we will formulate the parallel setup in QGEM; see  Fig.~\ref{fig:classicaltrajectoryHarmonic}.
Initially,  the wave function of two nanoparticles is defined as:
\begin{align}
    \ket{\psi(t=0)} = \frac{1}{2} (\ket{+}+\ket{-})_1 \otimes (\ket{-}+\ket{+})_2
\end{align}
%%%%%%%%%%%%%%%%%%%%%%%%%%%%%%%%
%%%%%%%%%%%%%%%%%%%%%%%%%%%%%%%%%
where $\ket{\pm}_k$ represents the two spatially separated paths of the matter-wave interferometer $k=\{1,2\}$. During the interferometer sequences, we have $\ket{ij}=\ket{i}_1 \ket{j}_2$, where $\{i,j\} \in \{+,-\}$, 
%both nanoparticles
pairs interact with each other due to the quantum nature of gravity and evolve according to the Hamiltonian $e^{-iH t/\hbar}$. 
Hence, the entanglement phase is determined by the gravitational potential energy, which dominates over their respective kinetic terms:
$$\phi_{ij} = -\frac{1}{\hbar}\int_0^\tau V[d_{ij}(t)] dt.$$
where $V$ is the total potential energy, $d_{ij}(t)$ is the distance between the pairs as shown in Fig.~\ref{fig:classicaltrajectoryHarmonic} and $\tau$ is the total interferometer time. During evolution, each path may accumulate additional phases due to the specific experimental setup. These are classical phases, such as between the nanoparticle and the electromagnetic shield, or classical disturbances between the nanoparticle and its local environment:
\begin{align}
    \ket{\psi(t)} = & \ \frac{1}{2}\Big( e^{i\phi_{++}}\ket{++}  + e^{i\phi_{+-}}\ket{+-} \nonumber\\
    & + e^{i\phi_{-+}}\ket{-+} + e^{i\phi_{--}}\ket{--} \Big)
\end{align}
%we set 
%\begin{equation}
%\phi^{(1)}_\pm + \phi^{(2)}_\pm = \phi_{\pm\pm}\,,~~~~~~~ \phi^{(1)}_\pm + \phi^{(2)}_\mp = \phi_{\pm \mp}\,,
%\end{equation}
%where $1,2$ correspond to the two matter-wave interferometers.
We assume that the spatial configuration is symmetric for states $\ket{++}$ and $\ket{--}$,  so that these two interaction phases become identical ($\phi_{++}=\phi_{--}=\phi$). 

Now, we can factorise and define the alternative phase 
\begin{equation}
\Delta\phi_+ = \phi_{-+}-\phi\,,~~~~\Delta\phi_{-} = \phi_{+-}-\phi.
\end{equation}
At the end of the experiment, the two trajectories of the two matter-wave interferometers will recombine, and the combined state becomes non-separable and entangled by virtue of the quantum gravitational interaction between the two interferometers; see~\cite{Bose:2017nin,Bose:2022uxe,Marshman:2019sne}. Hence, the final wavefunction of the combined state becomes:
\begin{align}
    \ket{\psi(t=\tau)} = \ &\frac{e^{i\phi}}{2}\Big(  \ket{++} + e^{i\Delta\phi_-} \ket{+-} \nonumber\\
    & +e^{i\Delta\phi_+}\ket{-+} + \ket{--}  \Big), \label{eq:evolution-of-wavefunction-with-generalize-phi}
\end{align}

We can construct the density matrix:
\begin{equation}
\rho = \ket{\psi}\bra{\psi}
\end{equation}
explicitly in Appendix~\ref{app:ppt-identical}, see Eq.~\eqref{eq:rho-pure}. For the difference in phases, we define the {\it entanglement phase}:
\begin{equation}
\phi_{ent}=\Delta\phi_- +\Delta\phi_+=\phi_{+-}+\phi_{-+}-2\phi\,.
\end{equation}
We can compute the entanglement witness $W$, defined by the Positive Partial Transpose (PPT) criterion~\cite{Horodecki:2009zz,Chevalier:2020uvv}. In the PPT criterion, if we partially transpose one of the subsystems (for example, subsystem 2) and the matrix has a negative eigenvalue, then the state can be called entangled. Performing the partial transpose (see Appendix~\ref{app:ppt-identical}), the eigenvalues are
%{\bf now use $\phi_{ent}$ directly to compactify the notation}

\begin{align}
    \lambda_{1,2} &= \pm\frac{1}{2} \sin{\left(\frac{\Delta\phi_++\Delta\phi_-}{2}\right)}, \label{eq:eig-pure-12}\\
    \lambda_{3,4} &= \frac{1}{2}\pm\frac{1}{2}\cos{\left(\frac{\Delta\phi_++\Delta\phi_-}{2}\right)} \label{eq:eig-pure-34}
\end{align}
This is a pure form of the eigenvalue. In real experiments, the interferometry system is an open quantum system experiencing environmental effects which are not just classical but also quantum decoherence~\cite{Schlosshauer:2019ewh,bassireview}, such as gas collisions, photons interacting with the nanoparticle, and the electromagnetic interactions between the nanoparticle and the chip that is levitating the nanoparticle, or the electromagnetic fluctuations between the shield, which is made up of a superconducting device. Now, whatever the source of decoherence, we add its effect over experimental time $t$ is determined by the accumulated dephasing:  
\begin{equation}
\Gamma = \gamma t,\label{eq:decoherence_rate}
\end{equation}
where $\gamma$ is the total decoherence rate, following Ref. \cite{Schut:2025blz}:
\begin{align}
    \bra{ij}\rho\ket{i'j'}\rightarrow e^{-\Gamma(2-\delta_{ii'}-\delta_{jj'})} \bra{ij}\rho\ket{i'j'}
\end{align}
where $\delta_{ii'}(\delta_{jj'})$ are the Kronecker delta function. The density matrix after the transpose is given in Eq.~\eqref{eq:rhoT2-gamma} of Appendix~\ref{app:ppt-identical}, and the eigenvalues are
\begin{align}
    \lambda_{1,2} &=\frac{1}{4}- \frac{e^{-\Gamma}}{4}\left[e^{-\Gamma} \pm 2 \sin{\left(\frac{\Delta\phi_+ +\Delta\phi_-}{2}\right)}\right], \label{eq:eig-gamma-12}\\
    \lambda_{3,4} &= \frac{1}{4}+ \frac{e^{-\Gamma}}{4}\left[e^{-\Gamma} \mp 2 \cos{\left(\frac{\Delta\phi_+ +\Delta\phi_-}{2}\right)}\right] \label{eq:eig-gamma-34}
\end{align}
\begin{figure*}
    \centering
    \includegraphics[width=0.95\linewidth]{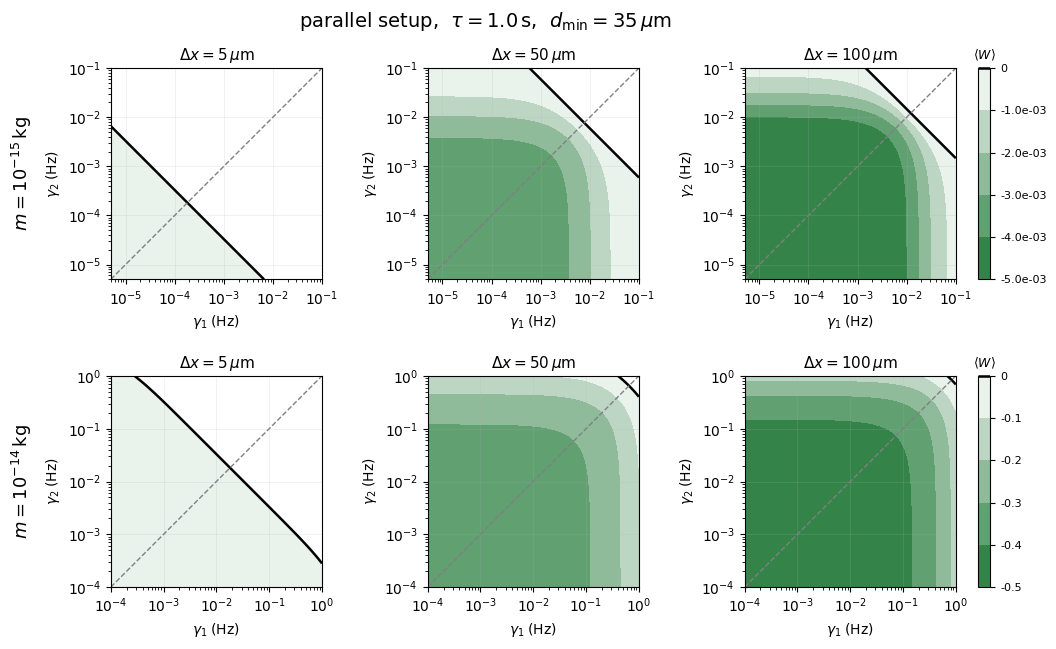}
    \caption{Parameter scan for the parallel 2-qubit configuration for 
    $\tau=1$ s and $d_\text{min} = 35\ \mu$m, for the superposition sizes $\Delta x=5,~50,~100{\rm \ \mu m}$ as a function of $\gamma_1,~\gamma_2$. The colour gradients show the PPT witness $\langle W \rangle < 0$ for the two qubits to get entangled solely via the quantum gravitational interaction (by a virtual graviton exchange); the black line demarcates the region where $\langle W \rangle = 0$. The white region shows no detection of entanglement under the PPT criteria for witnessing entanglement.   
    We select $d_{\rm min}=35 {\rm \mu m}$, which is based on the analysis that gravity between the two interferometers will dominate over Casimir, electric and magnetic dipole interactions between the nanoparticle and the conducting plate; see~\cite{Elahi:2024dbb}.
    The top panels correspond to $m = 10^{-15}$ kg and the bottom panels to $m = 10^{-14}$ kg. 
    The vertical dotted line shows where $\gamma_1=\gamma_2$. We can see that the witness is symmetrical around the line $\gamma_1=\gamma_2$. However, in these plots, it is easy to envisage a scenario in which $\gamma_1\neq \gamma_2$, and we can still maximise the witness by considering asymmetric scenarios such as either $\gamma_1> \gamma_2$ or $\gamma_2> \gamma_1$. This is the best possible scenario for detecting the gravitational entanglement for the chosen masses and the coherent splitting of the superposition.
    }
    \label{fig:parallel}
\end{figure*}

\begin{figure*}
    \centering
    \includegraphics[width=0.95\linewidth]{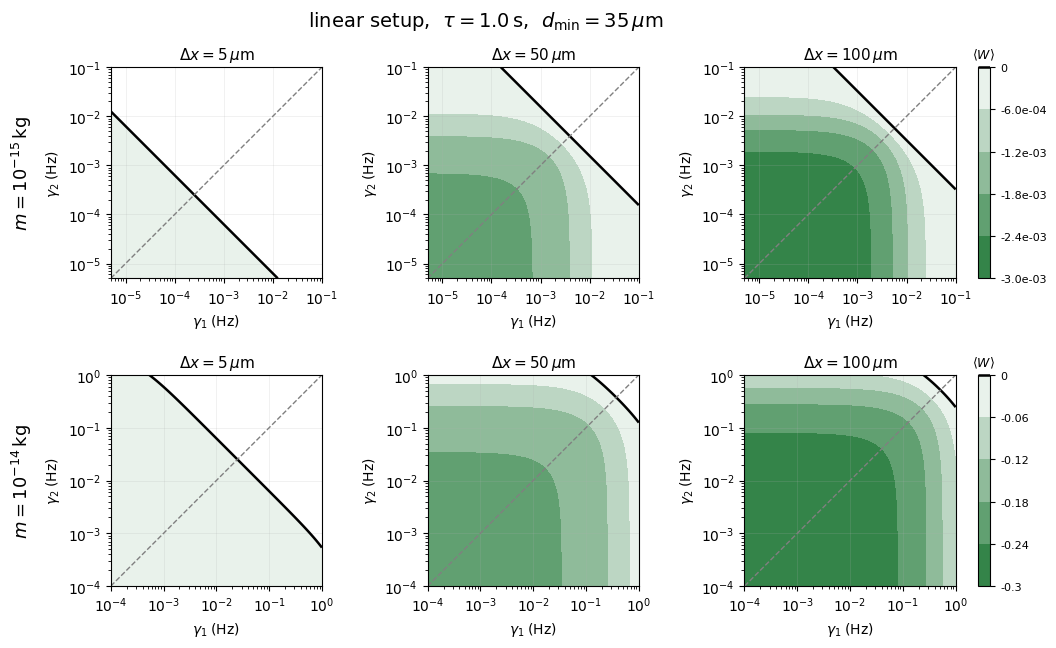}
    \caption{Parameter scan for the linear 2-qubit configuration for $\tau=1$ s and $d_\text{min} = 35\ \mu$m, for the superposition sizes $\Delta x=5,~50,~100{\rm \ \mu m}$ as a function of $\gamma_1,~\gamma_2$.
    The colour gradients show the PPT witness $\langle W \rangle < 0$ for the two qubits to get entangled solely via the quantum gravitational interaction (by a virtual graviton exchange); the black line demarcates the region where $\langle W \rangle = 0$. The dashed line shows $\gamma_1=\gamma_2$. 
The white region shows no detection of entanglement under the PPT criteria for witnessing entanglement. 
    We consider two different masses as shown above. We select $d_{\rm min}=35 {\rm \mu m}$, which is based on the analysis that gravity between the two interferometers will dominate over Casimir, electric and magnetic dipole interactions between the nanoparticle and the conducting plate; see~\cite{Elahi:2024dbb}. The trend is clear,
     in particular, the parallel configuration generally allows entanglement to be witnessed at smaller superposition sizes as compared to the linear case; see Fig.~\ref{fig:parallel}.}
    \label{fig:linear}
\end{figure*}

\begin{figure*}
    \centering
    \includegraphics[width=0.95\linewidth]{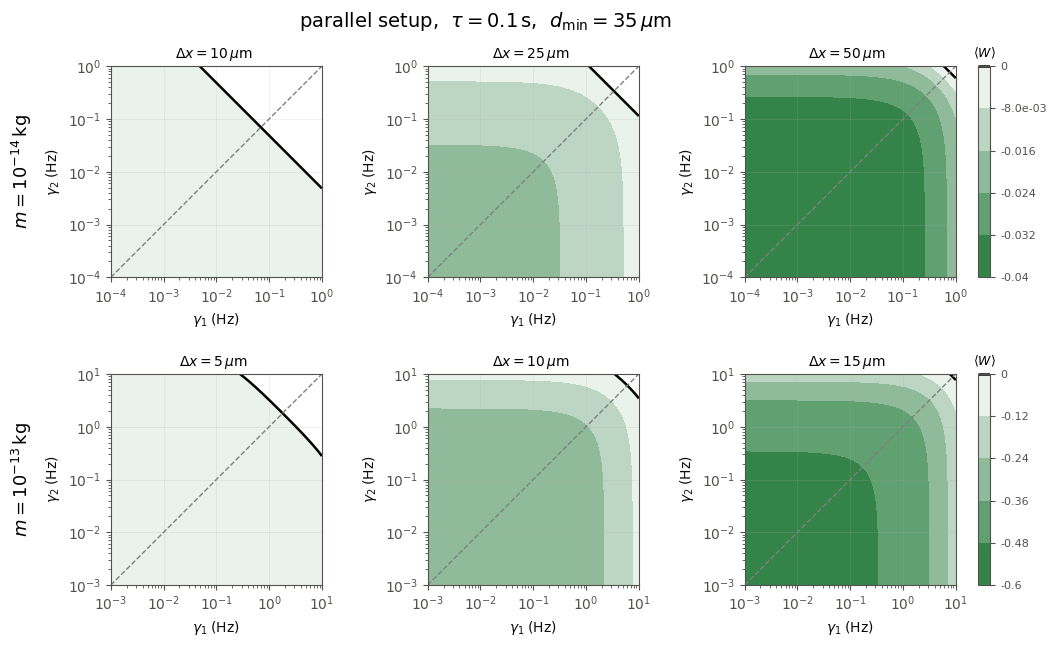}
    \caption{Parameter scan for the parallel 2-qubit configuration for time $\tau = 0.1$s, with $d_\text{min} = 35\ \mu$m, for the superposition sizes $\Delta x=10,~25,~50{\rm \ \mu m}$ for $m=10^{-14}$ kg and $\Delta x=5,~10,~15{\rm \ \mu m}$ for $m=10^{-13}$ kg  with respect to different values of $\gamma_1,~\gamma_2$.    
    The colour gradients show the PPT witness $\langle W \rangle < 0$ for the two qubits to get entangled solely via the quantum gravitational interaction (by a virtual graviton exchange); the black line demarcates the region where $\langle W \rangle = 0$.
   The dashed line shows $\gamma_1=\gamma_2$. 
    The colour gradients show the PPT witness $\langle W \rangle < 0$ for the two qubits to get entangled solely via the quantum gravitational interaction (by a virtual graviton exchange); the black line demarcates the region where $\langle W \rangle = 0$. The white region shows no detection of entanglement under the PPT criteria for witnessing entanglement. 
   Although we increased the mass by one order of magnitude, we checked that $d_{\rm min}=35~{\rm \mu m}$ remains suitable for placing the heavier mass in spatial superposition as well. Levitating an order-of-magnitude heavier nanodiamond will require changes to the current and magnetic-field configurations, but this does not impose additional constraints on the experiment.
    The shorter interaction time reduces the entanglement phase in general; hence, to overcome that, we moved to an $ m=10^{- 13} $~kg nanodiamond.
    Here we observe that $m=10^{-14}$~kg remains favourable over a broader range of decoherence rates for $\tau=1$~s experiment. However, note that to obtain a better witness, we must increase the mass by one order of magnitude, since the experimental time is shorter, i.e., $ \tau=0.1$ s.
    We can see that the witness improves remarkably.}
    \label{fig:parallel 0.1s}
\end{figure*}

\begin{figure*}
    \centering
    \includegraphics[width=0.95\linewidth]{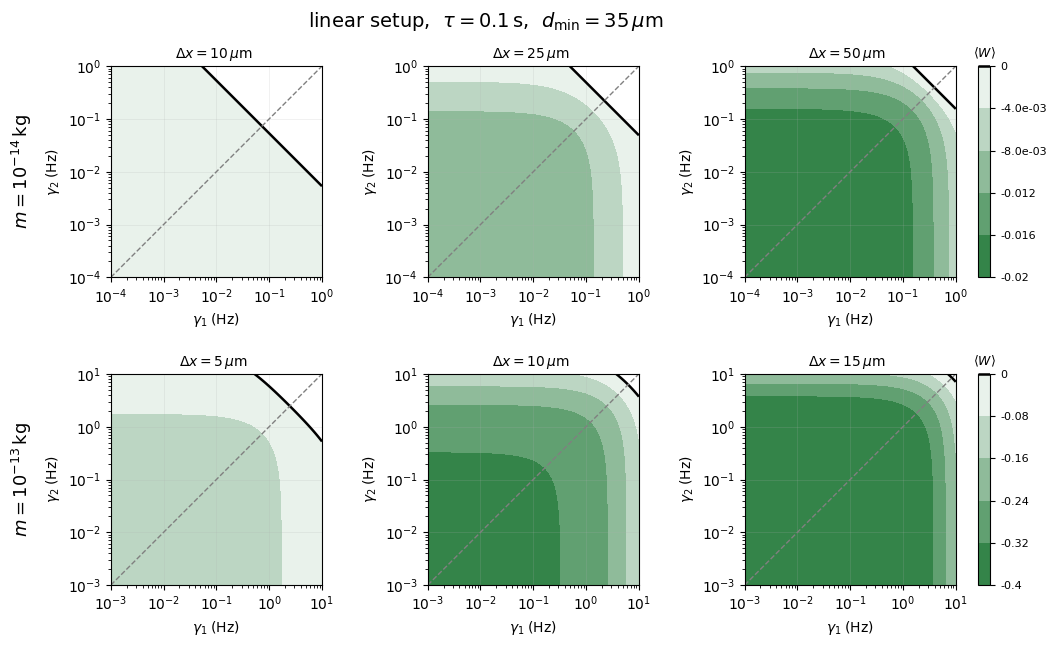}
    \caption{Parameter scan for the linear 2-qubit configuration at $\tau = 0.1$ s and $d_\text{min} = 35\ \mu$~m, for the superposition sizes $\Delta x=10,~25,~50{\rm \ \mu m}$ for $m=10^{-14}$ kg and $\Delta x=5,~10,~15{\rm \ \mu m}$ for $m=10^{-13}$ kg  with respect to different values of $\gamma_1,~\gamma_2$.
    We consider two different masses as shown above. The colour gradients show the PPT witness $\langle W \rangle < 0$ for the two qubits to get entangled solely via the quantum gravitational interaction (by a virtual graviton exchange); the black line demarcates the region where $\langle W \rangle = 0$. The white region shows no detection of entanglement under the PPT criteria for witnessing entanglement. The dashed line shows $\gamma_1=\gamma_2$.    
    We select $d_{\rm min}=35 {\rm \mu m}$, which is similar to the earlier cases. 
    Although we increased the mass by one order of magnitude, we checked that $d_{\rm min}=35 {\rm \mu m}$ remains suitable for placing the heavier mass in spatial superposition as well. Levitating an order-of-magnitude heavier nanodiamond will require changes to the current and magnetic-field configurations, but this does not impose additional constraints on the experiment.
    The shorter interaction time reduces the entanglement phase in general; hence, to overcome that, we moved to $ m=10^{- 13} $ kg nanodiamond, and we see that the witness improves remarkably, albeit at the cost of creating superposition with a heavier mass. }
    \label{fig:linear 0.1s}
\end{figure*}
If we set $\Gamma = 0$, the eigenvalues will return to the pure state. The witness of entanglement $W$ can be defined as the most negative eigenvalue ($\lambda_-$) of the PPT~\cite{PeresPPT,Horodecki:2009zz,Chevalier:2020uvv}:
\begin{align}
    W &= \left(\ket{\lambda_-^{T_2}}\bra{\lambda_-^{T_2}}\right)^{T_2}\\
    \langle W \rangle &= Tr(W\rho)=Tr\left(\ket{\lambda_-^{T_2}}\bra{\lambda_-^{T_2}} \rho^{T_2}\right) = \lambda_-^{T_2}
\end{align}
for parallel setup, the value of $$\phi_{ent}=\Delta\phi_+ + \Delta\phi_-<0 $$ and the most negative eigenvalue is
\begin{align}
    \langle W \rangle_{par} =\frac{1}{4}- \frac{e^{-\Gamma}}{4}\left[e^{-\Gamma} - 2 \sin{\left(\frac{\Delta\phi_+ +\Delta\phi_-}{2}\right)}\right],
    \label{W-parallel}
\end{align}
and for a linear setup, $$\phi_{ent}=\Delta\phi_+ + \Delta\phi_->0,$$ so the suitable eigenvalue for the entanglement witness for this setup is
\begin{align}
    \langle W \rangle_{lin} =\frac{1}{4}- \frac{e^{-\Gamma}}{4}\left[e^{-\Gamma} + 2 \sin{\left(\frac{\Delta\phi_+ +\Delta\phi_-}{2}\right)}\right],
    \label{W-lin}
\end{align}
In the limit when $\Gamma=\gamma t \ll1$, and $(\Delta \phi_{+}+\Delta \phi_{-})/2 \ll 1$, the expression simplifies to:
\begin{align}
    \langle W \rangle \approx \frac{\Gamma}{2} \pm \frac{(\Delta \phi_{+}+\Delta \phi_{-})}{4}
\end{align}
However, we may not be in the above regime for all the parameter scans. Therefore, we will keep the analysis as generic as possible and study Eqs.~(\ref{W-parallel}, \ref{W-lin}). 

\subsection{Asymmetrical decoherence between two states: Common and Differential modes}
\label{sec:asymmetric}

In experimental practice, especially in configurations that employ two interferometers that are spatially separated, the assumption of a perfectly symmetric decoherence rate ($\gamma_{1} = \gamma_{2}=\gamma$) is almost impossible to satisfy. When two masses or particles are sent through different interferometers, each will interact with an environmental reservoir that is specific to it. Asymmetry in the dephasing rates ($\gamma_{1}$ and $\gamma_{2}$) will almost certainly arise, for example, due to imperfections in the trapping potential, currents, tiny differences in the magnetic field.  Recall the wave function with the asymmetric decoherence assumption, with $\ket{i,j}$ where ($\{i,j\}\rightarrow\{+,-\}$), the evolved wavefunction is given by (see Appendix~\ref{app:global-phase}):
\begin{widetext}
\begin{align}
    \ket{\psi(\tau)} = \frac{e^{i[\varphi_+^{(1)}+\varphi_+^{(2)}]}}{2}\left( e^{i\phi_{++}}\ket{++}+e^{-i\delta\varphi^{(2)}} e^{i\phi_{+-}}\ket{+-} + e^{-i\delta\varphi^{(1)}} e^{i\phi_{-+}}\ket{-+}+e^{-i(\delta\varphi^{(1)}+\delta\varphi^{(2)})} e^{i\phi_{--}}\ket{--} \right)
\end{align}
\end{widetext}
where $$\delta\varphi^{1(2)} = \varphi^{1(2)}_+-\varphi^{1(2)}_{-},$$ 
denoting the global phase of each test mass $$\varphi_i^{1(2)} = -\left[\int_0^\tau L^{1(2)}_i dt\right]/\hbar,$$ 
%{\bf: AM: here $i$ stands for what?, same as $i,j=+, -$?}
where $L_i$ is the Lagrangian of each test mass in branch $i=\{+,-\}$ of its spatial superposition when it interacts with its own environment, such as trap potential, Earth's gravity, Casimir-Polder, residual electromagnetic interactions, etc. The gravitational induced entanglement phase is just: $\phi_{ij} = -\left[\int_0^\tau V(d_{ij}(t)) dt\right]/\hbar.$
For now, we assume the two masses accumulate different global phases ($\varphi_i^{(1)} \neq \varphi_j^{(2)}$). The global-phase factors act separately according to their subsystems, so both will be arranged into the diagonal unitary $(U_1\otimes U_2)$, and we can separate the global phase of the two masses from the gravitational phases (see Appendix~\ref{app:global-phase}).
%: $\ket{\psi(\tau)} = (U_1\otimes U_2)\ket{\psi_{ent}}$,
%where $\ket{\psi_{ent}} = {1}/{2} \sum_{i,j=\pm} e^{i\phi_{ij}}\ket{i,j}$
%{\bf Again entanglement phase is different here compared to Eq.6, why?}. 
Now, let us introduce that the two test masses experience different decoherence rates, so the density matrix also undergoes independent dephasing in each half of the electromagnetic shielding
\begin{align}
\bra{ij}\rho\ket{i'j'}\rightarrow e^{-[\Gamma_1(1-\delta_{ii'})+\Gamma_2(1-\delta_{jj'})]}\bra{ij}\rho\ket{i'j'},
\end{align}
or
\begin{widetext}
\begin{align}
    \bra{ij}\rho\ket{i'j'} = \frac{1}{4}  e^{i[\varphi_i^{(1)}-\varphi_{i'}^{(1)}]}  e^{i[\varphi_j^{(2)}-\varphi_{j'}^{(2)}]} e^{i(\phi_{ij}-\phi_{i'j'})} e^{-[\Gamma_1(1-\delta_{ii'})+\Gamma_2(1-\delta_{jj'})]}.\label{eq:density_matrix_2decoherence-in-general}
\end{align}
\end{widetext}
As shown in Appendix~\ref{app:global-phase}, the density matrix can be factorized as $\rho = (U_1\otimes U_2)\widetilde{\rho}(U_1\otimes U_2)^\dagger$, therefore $\rho$ and $\widetilde{\rho}$ have a similar eigenvalue \cite{PeresPPT}, and for the PPT of $\rho$ is equivalent to 
\begin{align}
    \rho^{T_2} = (U_1 \otimes U_2^*) \widetilde{\rho}^{T_2} (U_1 \otimes U_2^*)^\dagger
\end{align}
Since ($U_1 \otimes U_2^*$) is also unitary, it can be said that $\rho^{T_2}$ has the same eigenvalues as $\widetilde{\rho}^{T_2}$. Thus, the PPT eigenvalues (or later the Witness) as a whole do not depend on any of the global phases $\varphi_\pm^{1(2)}$, regardless of whether they are the same or different between the two test masses. This reminds us of the basic nature of entanglement itself: it is a property of the quantum interaction between the two masses, not an individual global phase that describes the classical phase of each test mass in its respective local background.

Now, we can focus only on the entanglement phase, but note that
\begin{align}
    \widetilde{\rho}_{(ij,i'j')} = \frac{1}{4} e^{i(\phi_{ij}-\phi_{i'j'})}e^{{-\Gamma_1}(1-\delta_{ii'})}e^{-\Gamma_2(1-\delta_{jj'})}.
\end{align}
After partial transposition, the element $\bra{ij}\widetilde{\rho}^{\,T_2}\ket{i'j'} = \bra{ij'}\widetilde{\rho}\ket{i'j}$ carries the phase $\phi_{ij'}-\phi_{i'j}$, and gives
\begin{align}
    \phi_{+-}+\phi_{-+}-\phi_{++}-\phi_{--} = \phi_{ent}. \label{eq:phi-eff-invariant}
\end{align}
As shown in Appendix~\ref{app:gauge}, the PPT eigenvalues depend on the entanglement phases only through this combination. We will therefore refer to $\phi_{ent}$ as the entanglement phase.
Note that Eq. \eqref{eq:phi-eff-invariant} is a general form of Eq. \eqref{eq:evolution-of-wavefunction-with-generalize-phi} (In the earlier treatment, the symmetric relationship was maintained, i.e. $\phi_{++} = \phi_{--}=\phi$ as a effective phase). The combination in \eqref{eq:phi-eff-invariant} remains well defined even when $\phi_{++} \neq \phi_{--}$~\footnote{
For future reference, whenever $\phi_{++} \neq \phi_{--}$ appears, it indicates a geometric mismatch between the two test masses. For example, the two superposition widths are unequal
($\Delta x_1 \neq \Delta x_2$, caused by mismatch in magnetic field gradients), or the two superposition axes are tilted with respect to each other~\cite{Schut:2023eux}, or the centre of one superposition is
laterally displaced, or different masses for both test masses.}.

To simplify, we introduce effective accumulated decoherence as a common mode of accumulated decoherence: 
\begin{equation}
\Gamma_{c} = \frac{1}{2} (\Gamma_1+\Gamma_2)
\end{equation}
and the asymmetry parameter as a differential mode of accumulated decoherence
\footnote{Note that $\Gamma_1$ and $\Gamma_2$ represent the accumulated decoherence phases associated with the environmental interactions experienced by each test mass (1 and 2), including gas molecule collision \cite{Schut:2024lgp}, blackbody radiation \cite{Romero-Isart:2011yun}, spin coherence of the NV-center \cite{WoodPRB22_GM}, jitter and gravity-gradient noise \cite{toros.PRR2021}, electromagnetic dipole-induced decoherence and trap-field noise \cite{Fragolino:2023agd}.}
\begin{equation}
\Gamma_{d} = \Gamma_1-\Gamma_2\,,
\end{equation}
we have $\Gamma_{1,2} = \Gamma_{c} \pm \Gamma_d/2$. 

The resulting density matrix $\widetilde{\rho}$ is given in Eq.~\eqref{eq:rho-tilde-raw} of Appendix~\ref{app:gauge}, and after PPT, we have the eigenvalues:
\begin{small}
\begin{align}
    \lambda_{1,2} &=\frac{1}{4}-\frac{e^{-\Gamma_{c}}}{4}\left[e^{-\Gamma_{c}}\pm\sqrt{2}\sqrt{\cosh{(\Gamma_d)}-\cos{(\phi_{ent})}}\right],\label{eq:asymmetrical_Eigen_Value_12} \\
    \lambda_{3,4} &=\frac{1}{4}+\frac{e^{-\Gamma_{c}}}{4}\left[e^{-\Gamma_{c}}\mp\sqrt{2}\sqrt{\cosh{(\Gamma_d)}+\cos{(\phi_{ent})}}\right], \label{eq:asymmetrical_Eigen_Value_34}
\end{align}
\end{small}
One can see that if  $\gamma_1=\gamma_2$, then the eigenvalues will return to Eqs.~\eqref{eq:eig-gamma-12} and \eqref{eq:eig-gamma-34}. Because the square root of Eq. \eqref{eq:asymmetrical_Eigen_Value_12} is non-negative, the most negative eigenvalue is the $+$ term, because $\cos{(\phi_{ent})} >0$ is always positive regardless of the sign of $\phi_{ent}$. Therefore, the PPT witnesses for both parallel and linear setups are represented by a single equation:
\begin{small}
\begin{align}
    \langle W\rangle &= \frac{1}{4}-\frac{e^{-\Gamma_{c}}}{4}\left[e^{-\Gamma_{c}} + \sqrt{2}\sqrt{\cosh{(\Gamma_d)}-\cos{(\phi_{ent})}}\right].
\end{align}    
\end{small}
Meanwhile, $\Gamma_c$ and $\Gamma_d$ have a physical constraint, i.e. $$2\Gamma_{c} \geq\abs{\Gamma_{d}}.$$ This is a consequence of the non-negative value of the decoherence parameter, so $\Gamma_1$ and $\Gamma_2$ fulfil the condition 
$\{\Gamma_1,\Gamma_2\} \geq 0$ so that $-2\Gamma_c\leq\Gamma_d\leq2\Gamma_c$.
In the small regime of phases $\{\Gamma_c,~\Gamma_d\} \ll 1 $ and $\phi_{ent} \ll 1$ we have 
\begin{align}
    \langle W\rangle \approx \frac{\Gamma_c}{2} - \frac{\sqrt{\Gamma_d^2 + \phi_{ent}^2}}{4} \label{eq:minimum_W}
\end{align}
where $\phi_{ent} > 2 \sqrt{\Gamma_c^2 - \Gamma^2_d/4}$.
Now we evaluate the entanglement phase, $\phi_{ent}$, first we set the positions of four arms (two belong to each mass (1) and (2)), and the distances between them are $d_{++}, d_{--}, d_{+-}$ and $d_{-+}$, see (Fig.~\ref{fig:classicaltrajectoryHarmonic}). The gravitational phase for each interaction is given by: $$\phi_{ij}=-\frac{1}{\hbar}\int_0^t dt V[d_{ij}(t)].$$
Recall the definition of an effective entanglement phase, again, $\phi_{ent} = {(\phi_{+-} + \phi_{-+})} - {(\phi_{++} + \phi_{--})}$ and 
%\begin{align}
%    \phi_{ent} = {(\phi_{+-} + \phi_{-+})} - {(\phi_{++} + \phi_{--})}\,. \nonumber
%\end{align}
let us now consider the two setups:

\begin{enumerate}
\item{Parallel setup}: In the parallel setup, we have: $d_{++} = d_{--} = d_{min}$ and $d_{+-}=d_{-+}=\sqrt{d^2_{min}+\Delta x^2}$, hence, we have
\begin{align}
    \phi_{ent}^{(par)} = 2\phi\left(\sqrt{d^2_{min}+\Delta x^2}\right)-2\phi(d_{min}),
\end{align}

\item{Linear setup}: For the linear setup, we have: $d_{++} = d_{--} = d_{min}+\Delta x$, $d_{+-}={d_{min}+2\Delta x}$ and $d_{-+}=d_{min}$, so that
\begin{small}
\begin{align}
    \phi_{ent}^{(lin)} = \phi({d_{min}+2\Delta x})+\phi(d_{min})-2\phi({d_{min}+\Delta x})\,.
\end{align}
\end{small}

\end{enumerate}

Although the creation of superposition is time-dependent, the largest entanglement phase for both parallel and linear setups is given by the case when 
the superposition size is the largest, and the largest phase accumulated during the experimental time scale is given by~\footnote{Strictly speaking, $\Delta x$ will have time dependence; just to illustrate, for $\Delta x < d_{\rm min}$ the entanglement phase is $\propto (\Delta x/d_{\rm min})^2$. For small time scales, we have  
$\phi_{ent}^{(par)}\approx -[Gm^2\tau/\hbar d_{\rm min}] [\Delta x/d_{\rm min}]^2$, while 
$\phi_{ent}^{(lin)}\approx [Gm^2\tau/\hbar d_{\rm min}] [\Delta x/d_{\rm min}]^2$. As we can see, the largest spatial superposition provides the best entanglement phase and the best witness, as well.
}: 
\begin{small}
\begin{align}
    \phi_{ent}^{(par)} &= \frac{2Gm^2}{\hbar}\left(\frac{1}{\sqrt{d^2_{min}+\Delta x^2}}- \frac{1}{d_{min}}\right)\tau, \label{eq:phi_parallel} \\
    \phi_{ent}^{(lin)} &= \frac{Gm^2}{\hbar}\left(\frac{1}{d_{min}+2\Delta x}+\frac{1}{d_{min}}- \frac{2} {d_{min}+\Delta x}\right) \tau, \label{eq:phi_linear}
\end{align}    
\end{small}
which are consistent with Ref. \cite{Schut:2025blz}. Note that the entanglement phase values for the parallel setup are always $\phi_{ent}^{(par)}<0$, and for the linear setup are $\phi_{ent}^{(lin)}>0$.

%%%%%%%%%%%%%%%%%%%%%%%%%%%%

\section{Discussion and Conclusion}
\label{sec:conclusion}

We have conducted an analytical calculation for the general case in which the two interferometers experience different decoherence rates, $\gamma_1$ and $\gamma_2$, and different global phases. Following the parameters in \cite{Schut:2025blz}, where the masses are $m=10^{-14}$ and $m=10^{-15}$kg and considering parallel and linear setups, we introduce the differential mode of dephasing, $\Gamma_d = \Gamma_1 -\Gamma_2$.

Figs. \ref{fig:parallel} and \ref{fig:linear} show that for small phases (entanglement and dephasing), $\langle W \rangle <0$ reduces to the condition $\sqrt{\gamma_1\gamma_2} < \phi_{ent}/(2\tau)$.
Since $\phi_{ent} \propto m^2$, the witness gradient for the two rows look the same at $\Delta x = 5 \ \mu m$, but for $m = 10^{-14}$ kg and $\Delta x \geq 50 \ \mu m$ the entanglement phase reaches $ \phi_{ent}\sim 1.5-2.4$ rad, the boundary (shown by the black line, and the minimum value of the witness is given by: $\langle W_{min}\rangle = -\frac{1}{2}\sin(\phi_{ent}/2)$, which approaches $\langle W_{min} \rangle \sim -0.5$. The witness for the linear setup is slightly better at $\Delta x = 5 \ \mu m$, whereas the parallel setup is twice as favourable at $\Delta x= 50$ and $100 \ \mu m$. For realistic decoherence rates of $\gamma_{1,2}\sim 10^{-2} - 10^{-1}$ Hz, $m = 10^{-15}$ kg is not favourable at $\tau = 1$ s, while $m = 10^{-14} $ kg with $\Delta x \geq 50 \ \mu m$ gives better entanglement witness.

Furthermore, we also considered the shorter time for the experiment $\tau=0.1$ s in Figs. \ref{fig:parallel 0.1s} and \ref{fig:linear 0.1s}, which has the same pattern as $\tau = 1$ s. Since the entanglement phase $\phi_{ent} \propto m^2\tau$, this makes the most negative witness drop from about $\langle W\rangle\sim -0.5 \Longrightarrow -0.04$ for $m=10^{-14}$~kg nanodiamonds. For $\tau = 0.1$ s and $m = 10^{-14}$ kg, the witness remains bad even when we increase the superposition size $\Delta x$ (See Figures~\ref{fig:parallel 0.1s} and~\ref{fig:linear 0.1s}). The reason can be seen from Eqs.~\eqref{eq:phi_parallel}-\eqref{eq:phi_linear}: the entanglement phase saturates at $2Gm^2\tau/(\hbar d_{\min}) \approx 0.36$~rad, far below the optimal value $\phi_{ent} = \pi$, so that $\langle W\rangle \lesssim 0.04$ regardless of $\Delta x$. According to the same equations, the phase can be increased by reducing $d_{min}$ or by increasing the mass. Increasing the mass by a factor of ten yields a factor of $10^2$ increase in the entanglement phase, which can only be matched by reducing $d_{\min}$ by a factor of $10^2$. Since $d_{\min} = 35~\mu$m is already fixed by the trap screen geometry \cite{Elahi:2024dbb}, for which gravity still dominates over Casimir-Polder, dipole-dipole and magnetic-moment contributions (between the nanodiamond and the superconducting screen, and between the nanodiamonds as far as gravity is concerned), the remaining option is to increase the mass to $m = 10^{-13}$ kg.

To improve the witness, we are then compelled to make the mass of the nanodiamond heavier, and this results in a vast improvement in the witness, which can be observed in Fig.(\ref{fig:parallel 0.1s} and~\ref{fig:linear 0.1s}). In fact, $\langle W\rangle\sim -0.2$ can be reached for the superposition size of $\Delta x=5~{\rm \mu m}$, and $\langle W\rangle\sim -0.4$ for $\Delta x =15~{\rm \mu m}$, for both parallel and linear configurations.

In summary, in either the $ \tau=1$ or 
$ \tau=0.1$~s experiment, the parallel configuration offers a more favourable parameter space for $\gamma_1,~\gamma_2$ for witnessing the quantum gravity-induced entanglement. For the $\tau=0.1$~s experiment, it is desirable from the perspective of creating spatial superposition on a smaller time scale; hence, it falls within the frequency band of $10$~Hz. The latter requires slightly less demanding acceleration noise power spectral density compared to the $1$~s or $1$~Hz experiment~\cite{toros.PRR2021}. So, the most desirable scenario will be to perform the $0.1$~s experiment for $m=10^{-14}$~kg, but we have to accept a poor witness $\langle W\rangle \sim- 0.04$ for $m=10^{-14}$~kg; hence, more repetitions will be required to build the confidence level appreciably high. On the other hand, by increasing the mass by one order, the witness improves drastically with a modest coherence superposition size. Creating spatial superposition is still challenging for such a range of masses, and the effort is ongoing
from both theoretical and experimental sides.
Nonetheless, our results are very important for benchmarking experiments, because the analysis and plots suggest the limitations and the parameters we should be aiming towards testing the  
quantum nature of gravity in a lab in a model-independent fashion.

\begin{acknowledgments}
We thank Martine Schut for a valuable discussion.
R.R. is supported by Beasiswa Indonesia Bangkit and Lembaga Pengelola Dana Pendidikan (BIB LPDP) of the Ministry of Religious Affairs of Indonesia. T.Z. is supported by the China Scholarship Council (CSC). A.M.'s research is funded by the Gordon and Betty Moore Foundation through Grant GBMF12328, DOI 10.37807/GBMF12328. This material is based on work supported by the Alfred P. Sloan Foundation under Grant No. G-2023-21130.
\end{acknowledgments}

\bibliographystyle{apsrev4-2}
\bibliography{ref.bib}

% =====================================================================
\clearpage
\newpage
\onecolumngrid
\appendix

\section{PPT eigenvalues for identical decoherence}
\label{app:ppt-identical}

The density matrix without involving decoherence is $\rho = \ket{\psi}\bra{\psi}$ of the state \eqref{eq:evolution-of-wavefunction-with-generalize-phi} is
\begin{align}
    \rho = \frac{1}{4}\sum_{ij,i'j'}e^{i(\phi_{ij}-\phi_{i'j'})}\ket{ij}\bra{i'j'},
\end{align}
as the decoherence involves in our system: $\bra{ij}\rho\ket{i'j'}\rightarrow e^{-\Gamma(2-\delta_{ii'}-\delta_{jj'})} \bra{ij}\rho\ket{i'j'}$, as $\Delta\phi_+ = \phi_{-+}-\phi, \Delta\phi_- = \phi_{+-}-\phi$, and $\phi=\phi_{++}=\phi_{--}$, we have the matrix form
\begin{align}
    \rho = \frac{1}{4}
    \begin{pmatrix}
        1 & e^{-i\Delta\phi_-}e^{-\Gamma} & e^{-i\Delta\phi_+}e^{-\Gamma} & e^{-2\Gamma}   \\
        e^{i\Delta\phi_-}e^{-\Gamma} & 1 & e^{i(\Delta\phi_- - \Delta\phi_+)}e^{-2\Gamma} & e^{i\Delta\phi_-}e^{-\Gamma} \\
        e^{i\Delta\phi_+} & e^{-i(\Delta\phi_- - \Delta\phi_+)} & 1 &  e^{i\Delta\phi_+}\\
         e^{-2\Gamma}& e^{-i\Delta\phi_-}e^{-\Gamma} & e^{-i\Delta\phi_+}e^{-\Gamma} & 1
    \end{pmatrix} \label{eq:rho-pure}
\end{align}
Next, we used the PPT formalism, and we get the density matrix 
\begin{align}
    \rho^{T_2} = \frac{e^{-\Gamma}}{4}
    \begin{pmatrix}
        e^{\Gamma} & e^{i\Delta\phi_-} & e^{-i\Delta\phi_+} & e^{-\Gamma}e^{i(\Delta\phi_- - \Delta\phi_+)}   \\
        e^{-i\Delta\phi_-} & e^{\Gamma} & e^{-\Gamma} & e^{i\Delta\phi_-} \\
        e^{i\Delta\phi_+} & e^{-\Gamma} & e^{\Gamma} &  e^{-i\Delta\phi_+}\\
         e^{-\Gamma}e^{i(\Delta\phi_- - \Delta\phi_+)}& e^{-i\Delta\phi_-} & e^{i\Delta\phi_+} & e^{\Gamma}
    \end{pmatrix} \label{eq:rhoT2-gamma}
\end{align}
with the eigenvalues given in Eqs.~\eqref{eq:eig-gamma-12} and \eqref{eq:eig-gamma-34}:
\begin{align}
    \lambda_{1,2} &=\frac{1}{4}- \frac{e^{-\Gamma}}{4}\left[e^{-\Gamma} \pm 2 \sin{\left(\frac{\Delta\phi_+ +\Delta\phi_-}{2}\right)}\right],\\
    \lambda_{3,4} &= \frac{1}{4}+ \frac{e^{-\Gamma}}{4}\left[e^{-\Gamma} \mp 2 \cos{\left(\frac{\Delta\phi_+ +\Delta\phi_-}{2}\right)}\right].
\end{align}
for the pure state $\Gamma=0$, whose eigenvalues are given in Eqs.~\eqref{eq:eig-pure-12} and \eqref{eq:eig-pure-34}.
% ---------------------------------------------------------------------
\section{Factorization of the global phases of the two matter-wave interferometers and invariance of the PPT spectrum}
\label{app:global-phase}

Recall the initial condition of the wave function on two test masses:
\begin{align}
    \ket{\psi(0)}=\frac{1}{{2}}\left( \ket{++}+\ket{+-}+\ket{-+}+\ket{--} \right)
\end{align}
as time evolves,
\begin{align}
    \ket{\psi(t)}=U(t)\ket{\psi(0)},
\end{align}
where $U(t) = \exp{-iHt/\hbar}$. As we mentioned before, test masses 1 and 2 interact with their own environment differently, and this addition will be in the Hamiltonian along with entanglement:
\begin{align}
    H=H^{(1)}_g\otimes1+1\otimes H_g^{(2)}+H_{ent}, 
\end{align}
because the branch states are eigenstates of the single masswhere
\begin{align}
    H^{(1)}_g & = \sum_{i=\pm} L_i^{(1)}(\ket{i}\bra{i})_1=
    \begin{pmatrix}
       L_+^{(1)} & 0 \\
        0 & L_-^{(1)}
    \end{pmatrix}, \\
    H^{(2)}_g & = \sum_{j=\pm} L_j^{(2)}(\ket{j}\bra{j})_2=
    \begin{pmatrix}
        L_+^{(2)} & 0 \\
        0 & L_-^{(2)}
    \end{pmatrix}, \\
    H_{ent} &= \sum_{i,j=\pm} V_{ij}\ket{ij}\bra{ij}=
    \begin{pmatrix}
        V_{++} & 0 & 0 & 0\\
        0 & V_{+-} & 0 & 0\\
        0 & 0 & V_{-+} & 0 \\
        0 & 0 & 0 & V_{--}
    \end{pmatrix},
\end{align}
Using Taylor expansion
\footnote{
using the Taylor expansion $e^{X} = 1+X+{X^2}/{2!}+ ...=\sum_n X^n/(n!)$, by matching with the matrix structure in this case $X = diag(a,b)$, we have
\begin{align*}
    e^{X} &= 
    \begin{pmatrix}
        1&0\\
        0&1
    \end{pmatrix}
    +
    \begin{pmatrix}
        a&0
        \\
        0&b
    \end{pmatrix}
    +\frac{1}{2!}
    \begin{pmatrix}
        a^2&0\\
        0&b^2
    \end{pmatrix}
    +\frac{1}{3!}
    \begin{pmatrix}
        a^3&0\\
        0&b^3
    \end{pmatrix}+...\\
    &=\begin{pmatrix}
        1+a+\tfrac{a^2}{2!}+... & 0\\
        0 & 1+b+\tfrac{b^2}{2!}+...
    \end{pmatrix}
    \\
    &=\begin{pmatrix}e^{a}&0\\0&e^{b}\end{pmatrix}.
\end{align*}
}
we define $U(t) = [U_1\otimes U_2] U_{ent}$ where
\begin{align}
    U_1 \otimes U_2 &= \exp{-i[H^{(1)}_g\otimes1+1\otimes H_g^{(2)}]t/\hbar} =
    \begin{pmatrix}
        e^{i[\varphi_+^{(1)}+\varphi_+^{(2)}]} & 0 & 0 & 0\\
        0 & e^{i[\varphi_+^{(1)}+\varphi_-^{(2)}]} & 0 & 0\\
        0 & 0 & e^{i[\varphi_-^{(1)}+\varphi_+^{(2)}]} & 0 \\
        0 & 0 & 0 & e^{i[\varphi_-^{(1)}+\varphi_-^{(2)}]}
    \end{pmatrix}, \\
    U_{ent} &= \exp{-iH_{ent}t/\hbar} = 
    \begin{pmatrix}
        e^{i\phi_{++}} & 0 & 0 & 0\\
        0 & e^{i\phi_{+-}} & 0 & 0\\
        0 & 0 & e^{i\phi_{-+}} & 0 \\
        0 & 0 & 0 & e^{i\phi_{--}}
    \end{pmatrix},
\end{align}
where $\varphi_i^{1(2)} = -\int_0^\tau L^{1(2)}_i dt/\hbar$ define as global phase and $\phi_{ij} = -\int_0^\tau V_{ij}dt/\hbar$ as entanglement phase. The global-phase factors act separately according to its subsystem, so both will be arranged as diagonal unitary $U_1\otimes U_2$. The explicit form of the evolution of the wave function as
\begin{align}
    \ket{\psi(\tau)} = \frac{e^{i[\varphi_+^{(1)}+\varphi_+^{(2)}]}}{2}\left( e^{i\phi_{++}}\ket{++}+e^{-i\delta\varphi^{(2)}} e^{i\phi_{+-}}\ket{+-} + e^{-i\delta\varphi^{(1)}} e^{i\phi_{-+}}\ket{-+}+e^{-i(\delta\varphi^{(1)}+\delta\varphi^{(2)})} e^{i\phi_{+-}}\ket{--} \right)
\end{align}
where $\delta\varphi^{1(2)} = \varphi^{1(2)}_+-\varphi^{1(2)}_-$. Recall the density matrix with asymmetrical decoherence in Eq.~\eqref{eq:density_matrix_2decoherence-in-general}
\begin{align}
    \bra{ij}\rho\ket{i'j'} = \frac{1}{4}  e^{i[\varphi_i^{(1)}-\varphi_{i'}^{(1)}]}  e^{i[\varphi_j^{(2)}-\varphi_{j'}^{(2)}]} e^{i(\phi_{ij}-\phi_{i'j'})} e^{-[\Gamma_1(1-\delta_{ii'})+\Gamma_2(1-\delta_{jj'})]},
\end{align}
one can see, the diagonal elements are all 1/4 and we have six density of matrix elements:
\begin{align}
    \bra{++}\rho \ket{+-} &=\frac{e^{i\delta\varphi^{(2)}}}{4} e^{i(\phi_{++}-\phi_{+-})}e^{-\Gamma_2}, &  \bra{++}\rho \ket{-+} & =\frac{e^{i\delta\varphi^{(1)}}}{4} e^{i(\phi_{++}-\phi_{-+})}e^{-\Gamma_1}, \\
    \bra{++}\rho \ket{--} & =\frac{e^{i(\delta\varphi^{(1)}+\delta\varphi^{(2)})}}{4} e^{i(\phi_{++}-\phi_{--})} e^{-\Gamma_1} e^{-\Gamma_2}, & \bra{+-}\rho \ket{-+} & =\frac{e^{i(\delta\varphi^{(1)}-\delta\varphi^{(2)})}}{4} e^{i(\phi_{+-}-\phi_{-+})} e^{-\Gamma_1} e^{-\Gamma_2}, \\
    \bra{+-}\rho \ket{--} & =\frac{e^{i\delta\varphi^{(1)}}}{4} e^{i(\phi_{+-}-\phi_{--})}e^{-\Gamma_1}, & \bra{-+}\rho \ket{--} &=\frac{e^{i\delta\varphi^{(2)}}}{4} e^{i(\phi_{-+}-\phi_{--})}e^{-\Gamma_2},
\end{align}
%\begin{align}
%    \rho = \frac{1}{4}
%    \begin{pmatrix}
%        1& {e^{i\delta\varphi^{(2)}}} e^{i(\phi_{++}-\phi_{+-})}e^{-\Gamma_2} & {e^{i\delta\varphi^{(1)}}} e^{i(\phi_{++}-\phi_{-+})}e^{-\Gamma_1} & {e^{i(\delta\varphi^{(1)}+\delta\varphi^{(2)})}} e^{i(\phi_{++}-\phi_{--})} e^{-\Gamma_1} e^{-\Gamma_2}\\
%        {e^{-i\delta\varphi^{(2)}}} e^{i(\phi_{+-}-\phi_{++})}e^{-\Gamma_2} & 1& {e^{i(\delta\varphi^{(1)}-\delta\varphi^{(2)})}}e^{i(\phi_{+-}-\phi_{-+})} e^{-\Gamma_1} e^{-\Gamma_2} & {e^{i\delta\varphi^{(1)}}} e^{i(\phi_{+-}-\phi_{--})}e^{-\Gamma_1} \\
%        {e^{-i\delta\varphi^{(1)}}} e^{i(\phi_{-+}-\phi_{++})}e^{-\Gamma_1}& {e^{-i(\delta\varphi^{(1)}-\delta\varphi^{(2)})}}e^{i(\phi_{-+}-\phi_{+-})} e^{-\Gamma_1} e^{-\Gamma_2} & 1& {e^{i\delta\varphi^{(2)}}} e^{i(\phi_{-+}-\phi_{--})}e^{-\Gamma_2}\\
%        {e^{-i(\delta\varphi^{(1)}+\delta\varphi^{(2)})}} e^{i(\phi_{--}-\phi_{++})} e^{-\Gamma_1} e^{-\Gamma_2}& {e^{-i\delta\varphi^{(1)}}} e^{i(\phi_{--}-\phi_{+-})}e^{-\Gamma_1} & {e^{-i\delta\varphi^{(2)}}} e^{i(\phi_{--}-\phi_{-+})}e^{-\Gamma_2} & 1
%    \end{pmatrix}
%\end{align}
and the other six follow $\rho_{i'j',ij} = \rho^*_{ij,i'j'}$ . Note that every test mass 1 and 2, contributes to its respective global phase difference $\delta\varphi_{1(2)}$ together with its decoherence rate. We can factorise the density matrix to separate the global phase (in $U_{1(2)}$) and the entanglement phase with the decoherence rate (in $\widetilde{\rho}$) \cite{PeresPPT,HornJohnson2013}:
\begin{align}
    {\rho} = (U_1 \otimes U_2)\widetilde{\rho}(U_1 \otimes U_2)^\dagger, &  & \widetilde{\rho}_{(ij,i'j')} = \frac{1}{4} e^{i(\phi_{ij}-\phi_{i'j'})}e^{-\Gamma_1(1-\delta_{ii'})}e^{-\Gamma_2(1-\delta_{jj'})}
\end{align}
Because the global phases matrix is a unitary matrix [$(U_1 \otimes U_2)(U_1 \otimes U_2)^\dagger =1$] According to the linear algebra property, $\rho$ and
$\widetilde{\rho}$ are related by a similarity transformation so that both have the same eigenvalue
\footnote{
Suppose that we have matrix definition of matrix $A=U B U^\dagger$ where $U$ is the unitary matrix, $\lambda$ and $v$ are respectively the eigenvalue and eigenvector of matrix $A$:
\begin{align*}
    Av &= \lambda v, \\ 
    UBU^\dagger v &= \lambda v, \\
    (U^\dagger U) B (U^\dagger v)&=\lambda(U^\dagger v),\\
    B (U^\dagger v)&=\lambda(U^\dagger v),
\end{align*}
here we prove that $A$ and $B$ have a same eigenvalue.
}
. Next, we conduct PPT at $\rho$, $\bra{ij}\rho^{T_2}\ket{i'j'} = \bra{ij'}\rho\ket{i'j}$ we have
\begin{align}
    \bra{ij}\rho^{T_2}\ket{i'j'} = e^{i[\varphi_i^{(1)}-\varphi_{i'}^{(1)}]} e^{-i[\varphi_j^{(2)}-\varphi_{j'}^{(2)}]} \bra{ij}\widetilde{\rho}^{T_2}\ket{i'j'}
\end{align}
The arrangement of the global phases of the two matter-wave interferometers is exactly the diagonal element of the unitary $U_1 \otimes U_2^*$ (where $*$ denotes the complex conjugate and $U_2$ from the transpose exchange $j$ and $j'$), or the PPT of $\rho$, which is also equivalent to
\begin{align}
    \rho^{T_2} = (U_1 \otimes U_2^*) \widetilde{\rho}^{T_2} (U_1 \otimes U_2^*)^\dagger.
\end{align}
Note that because $U_1\otimes U_2^*$ is also unitary, $\rho^{T_2}$ and
$\widetilde{\rho}^{T_2}$ also related by a similarity transformation and therefore have the same eigenvalues, so the global phases are not included in the PPT witness, which depends only on $\phi_{ij}$ and $\Gamma_{1(2)}$.
% ---------------------------------------------------------------------
\section{Eigenvalues of the density matrix with different accumulated phase}
\label{app:gauge}

We now directly evaluate the PPT spectrum, keeping all four entanglement phases $\phi_{ij}$ unchanged. We now directly evaluate the PPT spectrum, keeping all four entanglement phases $\phi_{ij}$ unchanged. The matrix form of $\widetilde{\rho}$ is
\begin{align}
    \widetilde\rho = \frac{1}{4}
    \begin{pmatrix}
        1 &
        e^{-\Gamma_2} e^{i(\phi_{++}-\phi_{+-})} & e^{-\Gamma_1} e^{i(\phi_{++}-\phi_{-+})} & e^{-\Gamma_1}e^{-\Gamma_2} e^{i(\phi_{++}-\phi_{--})} \\
        e^{-\Gamma_2} e^{i(\phi_{+-}-\phi_{++})} & 1 & e^{-\Gamma_1}e^{-\Gamma_2} e^{i(\phi_{+-}-\phi_{-+})} & e^{-\Gamma_1} e^{i(\phi_{+-}-\phi_{--})} \\
        e^{-\Gamma_1} e^{i(\phi_{-+}-\phi_{++})} & e^{-\Gamma_1}e^{-\Gamma_2} e^{i(\phi_{-+}-\phi_{+-})} & 1 & e^{-\Gamma_2} e^{i(\phi_{-+}-\phi_{--})} \\
        e^{-\Gamma_1}e^{-\Gamma_2} e^{i(\phi_{--}-\phi_{++})} & e^{-\Gamma_1} e^{i(\phi_{--}-\phi_{+-})} & e^{-\Gamma_2} e^{i(\phi_{--}-\phi_{-+})} & 1
    \end{pmatrix},
    \label{eq:rho-tilde-raw}
\end{align}
and after PPT, we obtain:
\begin{align}
    \widetilde\rho^{T_2} = \frac{1}{4}
    \begin{pmatrix}
        1 & e^{-\Gamma_2} e^{i(\phi_{+-}-\phi_{++})} & e^{-\Gamma_1} e^{i(\phi_{++}-\phi_{-+})} & e^{-\Gamma_1}e^{-\Gamma_2} e^{i(\phi_{+-}-\phi_{-+})} \\
        e^{-\Gamma_2} e^{i(\phi_{++}-\phi_{+-})} &  1 & e^{-\Gamma_1}e^{-\Gamma_2} e^{i(\phi_{++}-\phi_{--})} & e^{-\Gamma_1} e^{i(\phi_{+-}-\phi_{--})} \\
        e^{-\Gamma_1} e^{i(\phi_{-+}-\phi_{++})} & e^{-\Gamma_1}e^{-\Gamma_2} e^{i(\phi_{--}-\phi_{++})} & 1 & c_2e^{i(\phi_{--}-\phi_{-+})} \\
        e^{-\Gamma_1}e^{-\Gamma_2}e^{i(\phi_{-+}-\phi_{+-})} & e^{-\Gamma_1} e^{i(\phi_{--}-\phi_{+-})} & e^{-\Gamma_2} e^{i(\phi_{-+}-\phi_{--})} & 1
    \end{pmatrix}.
    \label{eq:rho-tilde-PPT}
\end{align}
the eigenvalues of $\widetilde\rho^{\,T_2}$ become
\begin{align}
    \lambda_{1,2} &=\frac{1}{4}-\frac{e^{-\Gamma_{c}}}{4}\left[e^{-\Gamma_{c}}\pm\sqrt{2}\sqrt{\cosh{(\Gamma_d)}-\cos{(\phi_{ent}})}\right], \\
    \lambda_{3,4} &=\frac{1}{4}+\frac{e^{-\Gamma_{c}}}{4}\left[e^{-\Gamma_{c}}\pm\sqrt{2}\sqrt{\cosh{(\Gamma_d)}+\cos{(\phi_{ent}})}\right].
\end{align}
here, we introduce the mean accumulated dephasing $\Gamma_c = (\Gamma_1+\Gamma_2)/2$ and the dephasing mismatch $\Gamma_d = \Gamma_1-\Gamma_2$, and the entanglement phase as $\phi_{ent} = \phi_{+-}+\phi_{-+}-\phi_{++}-\phi_{--}$.
\end{document}